%% file: main.tex
\documentclass[conference]{IEEEtran}
\IEEEoverridecommandlockouts

\usepackage{cite}
\usepackage{url}
\usepackage{amsmath,amssymb,amsfonts,amsthm}
\usepackage{graphicx}
\usepackage{textcomp}
\usepackage{xcolor}
\usepackage[ruled,noend,nosemicolon]{algorithm2e}
\usepackage{booktabs}
\usepackage{tikz}
\usetikzlibrary{arrows.meta,positioning,fit,backgrounds,calc}

\begin{document}

\title{CheckCo: Checkpoint-Coordinated Load Balancing for Skewed Event Streams at Scale
\thanks{This paper addresses the same load-balancing problem studied in two prior conference papers~\cite{Nasir:2015:PBC:ICDE,Nasir:2016:WTC:ICDE} and contributes \textsc{CheckCo}, a new dynamic-oracle algorithm.}}

\author{\IEEEauthorblockN{Muhammad Anis Uddin Nasir}
\IEEEauthorblockA{\textit{King} \\
Stockholm, Sweden \\
anis.nasir@king.com}
\and
\IEEEauthorblockN{Reza Sadraei}
\IEEEauthorblockA{\textit{King} \\
Stockholm, Sweden \\
reza.sadraei@king.com}
}

\maketitle

\begin{abstract}
Event streams are rarely uniform: a few dominant keys can generate orders of magnitude more events than the rest, overloading the workers that receive them and capping throughput.
We present \textsc{CheckCo}, a checkpoint-coordinated load-balancing algorithm built on a dynamic oracle that estimates per-key frequencies from the live stream, with no prior statistics.
Workers count events per key locally; at each checkpoint barrier, the counts are merged into a partitioning function that divides each key into a number of splits roughly proportional to its estimated frequency and packs the splits onto workers, and this function routes the next checkpoint window.
We derive approximation bounds on the resulting load imbalance from classical scheduling results and bound its total key fan-out.
On a production pipeline processing over 100 billion events per day, the most loaded worker receives only $9.8\%$ more events than the average over a day (imbalance ratio $1.10$), against $8.03$ for key-based routing in simulation; in simulation, \textsc{CheckCo} also needs less than half the fan-out of earlier partitioners that reach comparable balance.
Shifts in the key distribution are reflected in routing from the next checkpoint on, and the load stays balanced through job restarts.
\end{abstract}


\input{sections/intro}
\input{sections/prelim}
\input{sections/problem}
\input{sections/algorithm}
\input{sections/analysis}
\input{sections/lakehouse}
\input{sections/experiments}
\input{sections/related}
\input{sections/conclusion}

\section*{Acknowledgment}
The authors thank Walter Chung and David Artiga for their contributions during the hackathon that originated this work.
Claude (Anthropic), used through Claude Code, was used extensively in writing this paper.
It drafted text in all sections, including the derivations in Section~\ref{sec:analysis}, checked the bibliography, and wrote the scripts that query the production metrics, run the simulations of Section~\ref{sec:eval} and generate the figures.
The authors reviewed and verified all AI-generated content.

\bibliographystyle{IEEEtran}
\bibliography{references}

\end{document}

%% file: sections/intro.tex
\section{Introduction}
\label{sec:intro}

Distributed stream processing engines~\cite{Carbone:2017:Flink} process continuous, high-throughput event streams by routing each event to a downstream operator subtask.
Each event carries a \emph{key}, such as its event type or the identifier of the entity that produced it.
A natural and widely-adopted routing strategy is to partition by key, co-locating all events of the same key at the same subtask.
This is required when downstream operators maintain per-key state, and is commonly applied even in stateless pipelines to simplify output organisation.

\begin{figure}[t]
\centering
\begin{tikzpicture}[
    >=Stealth,
    every node/.style={font=\scriptsize},
    worker/.style={draw, minimum width=7mm, anchor=south, fill=blue!25},
    workerOvr/.style={draw, minimum width=7mm, anchor=south, fill=red!60},
    lbl/.style={font=\tiny, align=center},
  ]

  \node[font=\scriptsize\bfseries] at (1.5,3.6) {(a) \textsc{KeyBy}};

  \node[workerOvr, minimum height=27mm] (wA) at (0.4,0)  {};
  \node[worker,    minimum height=10mm] (wB) at (1.1,0)  {};
  \node[worker,    minimum height=1mm]  (wC) at (1.8,0)  {};
  \node[worker,    minimum height=1mm]  (wD) at (2.5,0)  {};

  \node[lbl] at (0.4,2.3)  {\textbf{!}};
  \node[lbl, above=1pt of wA.north] {$w_1$};
  \node[lbl, above=1pt of wB.north] {$w_2$};
  \node[lbl, above=1pt of wC.north] {$w_3$};
  \node[lbl, above=1pt of wD.north] {$w_4$};

  \node[lbl, text=red!70!black]   at (0.4,-0.5) {$A$};
  \node[lbl] at (1.1,-0.5) {$B$};
  \node[lbl] at (1.8,-0.5) {$C$};
  \node[lbl] at (2.5,-0.5) {$D$};

  \draw[->] (0,0) -- (0,3.3) node[left,font=\tiny]{load};

  \node[font=\scriptsize\bfseries] at (5.5,3.6) {(b) \textsc{CheckCo}};

  \node[worker, minimum height=10mm] (pA) at (3.9,0) {};
  \node[worker, minimum height=10mm] (pB) at (4.6,0) {};
  \node[worker, minimum height=9mm]  (pC) at (5.3,0) {};
  \node[worker, minimum height=10mm] (pD) at (6.0,0) {};

  \node[lbl, above=1pt of pA.north] {$w_1$};
  \node[lbl, above=1pt of pB.north] {$w_2$};
  \node[lbl, above=1pt of pC.north] {$w_3$};
  \node[lbl, above=1pt of pD.north] {$w_4$};

  \node[lbl, text=red!70!black] at (3.9,-0.5) {$A_1,D$};
  \node[lbl, text=red!70!black] at (4.6,-0.5) {$A_2,C$};
  \node[lbl, text=red!70!black] at (5.3,-0.5) {$A_3$};
  \node[lbl] at (6.0,-0.5) {$B$};

  \draw[->] (3.5,0) -- (3.5,3.3) node[left,font=\tiny]{load};

  \draw[dashed, gray] (3.5,0.975) -- (6.3,0.975)
      node[right,font=\tiny,gray]{avg};

\end{tikzpicture}
\caption{Illustration of load imbalance due to skew.
(a)~\textsc{KeyBy} sends all events of the dominant key~$A$ to worker~$w_1$, overloading it (red bar) while $w_3$ and $w_4$ are nearly idle.
(b)~\textsc{CheckCo} splits key~$A$ across three workers, achieving near-uniform load.}
\label{fig:imbalance-intro}
\end{figure}
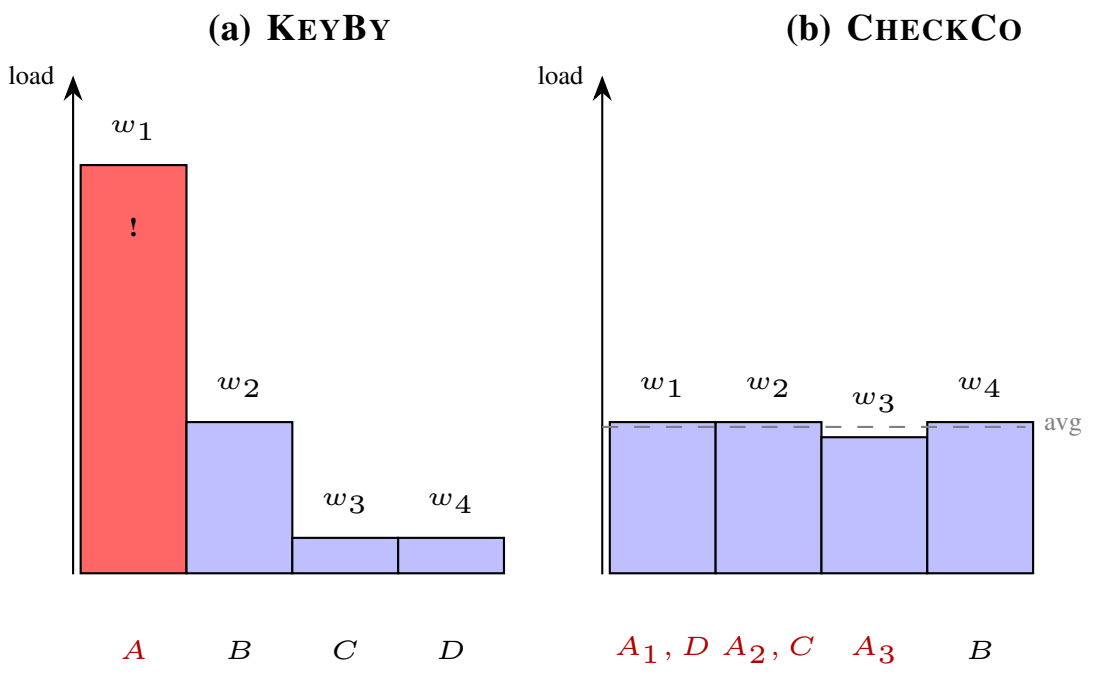

Under a uniform distribution of keys, key-based partitioning achieves good load balance.
In practice, however, key distributions are heavily skewed: a small number of keys account for the overwhelming majority of traffic, while most keys are rare.
This Zipf-like behaviour is common: it appears in web workloads~\cite{Breslau:1999:WCZ}, in the streams used to evaluate earlier partitioners~\cite{Gedik:2014,Nasir:2015:PBC:ICDE}, and in the game telemetry stream we study (Section~\ref{sec:eval}).
Routing all events of the same key to the same subtask (the standard \textsc{KeyBy} approach) concentrates the dominant keys on a small number of subtasks, leaving others idle, and the most loaded subtask then limits the throughput of the whole pipeline.
At the scale of production deployments processing over 100 billion events per day, even modest imbalance translates directly to wasted compute and lower throughput.
Figure~\ref{fig:imbalance-intro} illustrates the contrast.

Balancing a skewed stream requires sending the events of a heavy key to several workers.
We call the number of workers that receive a key its \emph{fan-out}.
Fan-out is the price of balance: every extra worker per key adds a shuffle channel and downstream state, and balance alone is trivial to reach by spreading every key over all workers, as shuffle grouping does.
The question is therefore how many workers each key needs, and prior solutions answer it without knowing the load of each key.

Some methods give heavy keys too few workers.
Key-grouping methods place every key on a single partition: Gedik~\cite{Gedik:2014} combines an explicit routing table for heavy keys with consistent hashing for the rest and tracks frequencies online, and Rivetti et al.~\cite{Rivetti:2015} learn the frequent keys in an initial phase and then deploy a fixed mapping.
PKG~\cite{Nasir:2015:PBC:ICDE} routes each event to the less loaded of two candidate workers.
A key that carries more than $d/P$ of the stream, with $d=1$ or $d=2$ respectively, overloads the workers that hold it however the other keys are placed; in the stream we study, the most frequent key carries $5.2\%$ of the events, more than the $2/P \approx 1.6\%$ that two of $P=128$ workers can absorb.
Other methods give heavy keys too many workers.
Our ICDE 2016 algorithms~\cite{Nasir:2016:WTC:ICDE} identify the heavy keys online and spread each of them over the same number of workers, all of them in the case of W-Choices, and FlexD~\cite{Liu:2024:FlexD} likewise expands the set of workers that serve high-frequency keys.
These methods balance well, but a key just above the heavy-hitter threshold receives as many workers as the heaviest key: on the stream we study, D-Choices gives each of the $106$ head keys $7$ workers, although $70$ of them fit on one, and W-Choices gives them all $128$, so the fan-out grows far beyond what balance requires.
Finally, methods that learn the key distribution once, such as Rivetti et al.~\cite{Rivetti:2015}, cannot follow it when it changes.

In this paper, we present \textsc{CheckCo}, a checkpoint-coordinated load-balancing algorithm that gives each key as many workers as its load requires.
We address the same problem as our earlier work~\cite{Nasir:2015:PBC:ICDE,Nasir:2016:WTC:ICDE}, but with an oracle that estimates the frequency of every key.
With these frequencies, \textsc{CheckCo} splits each key in advance into pieces of at most the average worker load, roughly in proportion to its frequency, and packs the pieces on the workers with the LPT heuristic.
Light keys stay on one worker and only heavy keys are split, so balance comes at close to the minimum fan-out.

The oracle is built from the live stream itself, with no external data source.
Each subtask sees only part of the stream, so the local counts must be merged, and routing should change only once the merge is complete.
Flink's checkpoint barriers provide such a point.
Each subtask counts keys between two checkpoints; at the barrier, the counts are merged and the partitioning function is recomputed, and the function built from one checkpoint window routes the next.
A shift in the key distribution is therefore reflected in routing from the next checkpoint on, and because the counts are kept in checkpointed state, they are also restored after a restart.

\paragraph{Contributions}
\begin{itemize}
  \item We show that per-key frequency estimates make the load-balancing problem of our earlier work easier to solve: splitting each heavy key in advance into pieces no larger than the target load turns it into classical multiprocessor scheduling, which we solve with Longest Processing Time (LPT) packing (\S\ref{sec:problem}, \S\ref{sec:algo}).
  Known scheduling bounds~\cite{Graham:1966:Bounds,Graham:1969:LPT} then carry over directly, while the fan-out stays at most $|\mathcal{K}|+P$ (\S\ref{sec:analysis}).
  \item We describe a dynamic oracle built on Flink's checkpoint mechanism, which merges the local per-key counts at every checkpoint barrier and keeps them in broadcast state, together with a sliding-window buffer that keeps the estimates stable during low-traffic periods (\S\ref{sec:dynamic}, \S\ref{sec:bootstrap}).
  \item We evaluate \textsc{CheckCo} on a production event ingestion pipeline processing over 100 billion events per day for a single streaming application (\S\ref{sec:eval}).
  Over a day, the most loaded worker receives $9.8\%$ more events than the average, against a mean imbalance ratio of $8.03$ estimated by simulation for key-based routing, and over 30 days the daily ratio stays between $1.04$ and $1.12$.
  In simulation on the same day, \textsc{CheckCo} stays within $5\%$ of the minimum fan-out, less than half that of D-Choices and an order of magnitude below W-Choices, which balance the stream slightly better, while two-choice routing cannot balance it.
  Shifts in the key distribution are reflected in routing from the next checkpoint on.
\end{itemize}

\paragraph{Roadmap}
Section~\ref{sec:prelim} introduces the stream-processing model and the oracle abstraction.
Section~\ref{sec:problem} states the load-balancing problem.
Section~\ref{sec:algo} describes \textsc{CheckCo}, including the dynamic oracle and bootstrapping strategies.
Section~\ref{sec:analysis} provides an approximation analysis, and Section~\ref{sec:lakehouse} discusses its implications for ingestion into a data lakehouse.
Section~\ref{sec:eval} presents the experimental evaluation.
Section~\ref{sec:related} surveys related work and Section~\ref{sec:conclusion} concludes.

%% file: sections/prelim.tex
\section{Preliminaries}
\label{sec:prelim}

\paragraph{Distributed stream processing}
We model a stream-processing application as a directed acyclic graph (DAG) of operators.
Each operator runs with a configurable \emph{parallelism} $P$: it is instantiated as $P$ concurrent subtasks, each processing a disjoint subset of the input records.
Records flow between operators over logical channels; the runtime implements each logical channel as $P \times P$ physical connections between subtask pairs.

We target Apache Flink~\cite{Carbone:2017:Flink}.
\textsc{CheckCo} relies on a small set of Flink concepts, which we define here and use throughout the paper.

\paragraph{Flink terminology}
\begin{itemize}
  \item \textbf{Subtask.} The unit of parallel execution: an operator with parallelism $P$ runs as $P$ subtasks, each with its own local state. Records are routed between subtasks by a \emph{partitioner}; \emph{key-based} partitioning sends all records with the same key to the same subtask, which is the source of the imbalance studied in this paper.

  \item \textbf{Checkpoint.} A periodic, globally consistent snapshot of the state of every operator, taken every checkpoint interval $\Delta_{\mathrm{ckpt}}$ and written to durable storage. On failure, the job restarts from the latest completed checkpoint, so each operator's state is restored to the value it held when that checkpoint was taken.

  \item \textbf{Checkpoint barrier.} A special marker that the sources inject into the stream to start a checkpoint. Barriers flow downstream with the records and separate those that belong to the current checkpoint from those that belong to the next. With aligned checkpoints, a subtask that has several inputs waits until it has received the barrier from all of them before it snapshots its state and forwards the barrier. The barrier therefore defines a consistent cut across all subtasks.

  \item \textbf{Checkpoint hooks.} Operators can run code at checkpoint time. \texttt{prepareSnapshotPreBarrier} is invoked on a subtask before the barrier is forwarded downstream, so records emitted from it reach the next operator ahead of the barrier. \texttt{snapshotState} is invoked when the subtask snapshots its state.

  \item \textbf{Operator state.} State that is local to a subtask and included in each checkpoint, as opposed to \emph{keyed state}, which is scoped to a record key.

  \item \textbf{Broadcast state.} A form of operator state in which a \emph{broadcast stream} is sent to every subtask of an operator, and each subtask stores an identical copy of the resulting key--value map~\cite{Flink:BroadcastState}. Only the broadcast side may update it; the regular input stream can read it. Because Flink has no cross-task communication, updates made on the broadcast side must be deterministic so that all copies stay identical. Each subtask checkpoints its own copy, which is kept in memory at runtime, and restores it on recovery. In a \texttt{BroadcastProcessFunction}, \texttt{processElement} handles regular records and \texttt{processBroadcastElement} handles broadcast records.

  \item \textbf{Side output.} An additional, typed output of an operator, separate from its main output. It lets an operator emit auxiliary records (here, local counts) without altering the main record stream.
\end{itemize}

\paragraph{Keys and terminology}
Let $S$ be a stream of events, each carrying a key drawn from a finite set $\mathcal{K}$.
Events are routed to $P$ downstream operator subtasks by a routing function $\pi$, which Section~\ref{sec:problem} defines formally together with the load-balancing problem.
Throughout the paper, \emph{workers}, \emph{subtasks} and \emph{partitions} all denote these $P$ parallel instances, and \emph{routing function} and \emph{partitioning function} both denote $\pi$.

\paragraph{The oracle abstraction}
An \emph{oracle} $\Omega$ is a component that supplies an estimate $\hat{c}(k)$ of the frequency of each key $k \in \mathcal{K}$.
A \emph{static oracle} provides $\hat{c}$ once at startup and never updates it.
A \emph{dynamic oracle} updates $\hat{c}$ periodically as traffic patterns evolve.
\textsc{CheckCo} uses a dynamic oracle: $\hat{c}$ is refreshed at every Flink checkpoint, and a static source may only seed it (Section~\ref{sec:bootstrap}).

%% file: sections/problem.tex
\section{Problem Statement}
\label{sec:problem}

We consider the load-balancing problem of our earlier work~\cite{Nasir:2015:PBC:ICDE,Nasir:2016:WTC:ICDE}.
The stream $S$ is a sequence of events $\langle t, k, v \rangle$ in timestamp order, where $t$ is the timestamp, $k \in \mathcal{K}$ the key and $v$ the value.
A partitioning function $\pi_t$ sends each event with timestamp $t$ to one of the $P$ workers according to its key.
Because the function is indexed by time, it may send events of the same key to different workers, as PKG does.

The load $\ell_p(t)$ of worker $p$ at time $t$ is the number of events with timestamp at most $t$ that were sent to $p$, and the imbalance at time $t$ is
\[
  I(t) = \max_p \ell_p(t) - \operatorname*{avg}_p \ell_p(t).
\]
The average load is the number of events up to $t$ divided by $P$ and does not depend on the partitioning function, so minimising the imbalance is the same as minimising the maximum load.
As in~\cite{Nasir:2015:PBC:ICDE}, we tackle the problem of finding a stream partitioning function that minimises the imbalance, while at the same time avoiding the downsides of shuffle grouping.

The \emph{fan-out} $\varphi_\pi(k)$ of a key $k$ is the number of distinct workers to which the partitioning function in force may send events of $k$: one for key grouping, two for PKG, and $P$ for shuffle grouping.
Shuffle grouping balances the load perfectly but gives every key fan-out $P$; every extra worker per key adds a shuffle channel and per-key state downstream.
Key grouping minimises fan-out but cannot balance a key that carries more than $1/P$ of the stream, and PKG cannot balance one that carries more than $2/P$.
Between these extremes, the question is how many workers each key needs.
Fan-out is therefore the cost of balancing a skewed stream; it is not the objective.

%% file: sections/algorithm.tex
\section{The \textsc{CheckCo} Algorithm}
\label{sec:algo}

Table~\ref{tab:notation} summarises the main notation used in this section.

\begin{table}[b]
  \centering
  \caption{Notation.}
  \label{tab:notation}
  \begin{tabular}{@{}l p{6.4cm}@{}}
    \toprule
    Symbol & Meaning \\
    \midrule
    $\mathcal{K}$ & Set of keys \\
    $P$ & Number of downstream partitions (parallelism) \\
    $\hat{c}(k)$ & Estimated frequency of key $k$ (oracle output) \\
    $N$ & Total estimated event count, $\sum_k \hat{c}(k)$ \\
    $L$ & Target load per partition, $\lceil N/P\rceil$ \\
    $\sigma(k)$ & Number of splits of key $k$ \\
    $\langle k,i\rangle$ & The $i$-th split of key $k$, $0\le i<\sigma(k)$ \\
    $s(\langle k,i\rangle)$ & Estimated load (size) of split $\langle k,i\rangle$ \\
    $\pi(k,i)$ & Partition assigned to split $\langle k,i\rangle$ (partitioning function) \\
    $\mathit{uid}(e)$ & Per-event non-negative integer identifier used for routing \\
    $W_x$, $\pi_x$ & Checkpoint window $x$; partitioning function built at barrier $x$ and applied to $W_{x+1}$ \\
    \bottomrule
  \end{tabular}
\end{table}

\subsection{Core: Split-Based Proportional Partitioning}
\label{sec:core}

\textsc{CheckCo} simplifies the problem of Section~\ref{sec:problem} in two steps: it first divides every key into pieces that fit on one partition, called \emph{splits}, whose number is determined by its estimated frequency (Equation~\eqref{eq:numsplits}), and then packs the splits onto the partitions with LPT.
Light keys keep a single split and heavy keys receive roughly as many splits as their load requires, which keeps the fan-out close to its minimum (Section~\ref{sec:analysis}).
The core algorithm takes as input a frequency map $\hat{c} : \mathcal{K} \to \mathbb{Z}_{\ge 0}$ of estimated counts and the number of downstream partitions $P$, and produces an assignment $\pi$ of splits to partitions.

\paragraph{Target load and split counts}
Define the total event count $N = \sum_{k \in \mathcal{K}} \hat{c}(k)$ and the target load per partition:
\[
  L = \left\lceil \frac{N}{P} \right\rceil.
\]
For each key $k$, we compute a split count:
\begin{equation}
  \label{eq:numsplits}
  \sigma(k) = \left\lceil \frac{\hat{c}(k)}{L} \right\rceil.
\end{equation}
Every key with $\hat{c}(k) \ge 1$ gets at least one split, a key with $\hat{c}(k) \le L$ keeps a single split, and since $\hat{c}(k) \le N \le PL$, no key gets more than $P$ splits.

\paragraph{Split enumeration and sizing}
For each key $k$ we create $\sigma(k)$ splits, labelled $\langle k, 0 \rangle, \langle k, 1 \rangle, \ldots, \langle k, \sigma(k)-1 \rangle$. The estimated size (load) of each split is:
\[
  s(\langle k, i \rangle) = \left\lceil \frac{\hat{c}(k)}{\sigma(k)} \right\rceil \le L.
\]
All splits of the same key carry equal estimated load, and no split exceeds the target load.

\paragraph{LPT bin-packing}
We assign splits to partitions using the Longest Processing Time (LPT) greedy algorithm~\cite{Graham:1969:LPT}.
Sort all splits in non-increasing order of $s(\cdot)$.
Maintain a min-heap $H$ of $P$ bins, each tracking its current total load.
For each split (in sorted order): pop the least-loaded bin from $H$, assign the split to it, update its load, and re-insert it.
Ties in bin load are broken by bin identifier, making the assignment deterministic across subtasks.

The result is the split counts $\sigma$ together with a map $\pi : \mathcal{K} \times [\sigma(\cdot)] \to [P]$ where $\pi(k, i)$ is the partition assigned to split $\langle k, i \rangle$; routing needs both (see below).

\paragraph{Hash routing}
At runtime, each incoming event $e$ of key $k$ is routed as follows.
Let $\mathit{uid}(e)$ be the unique identifier that every event carries, a non-negative integer.
The split index is:
\[
  i = \mathit{uid}(e) \bmod \sigma(k),
\]
and the destination partition is $\pi(k, i)$.
A key that is absent from $\hat{c}$, such as a key that first appears in the current window, is sent to partition $P-1$ until the next checkpoint.
This computation is stateless and requires no coordination between subtasks.
The time-indexed partitioning function $\pi_t$ of Section~\ref{sec:problem} is thus realised by the split map $\pi$ in force at time $t$, which is replaced at every checkpoint (Section~\ref{sec:dynamic}).

Algorithm~\ref{alg:core} summarises the full procedure.

\begin{algorithm}[t]
\caption{Split-Based Proportional Partitioning}
\label{alg:core}
\KwIn{Frequency map $\hat{c}$, number of partitions $P$}
\KwOut{Split counts $\sigma : \mathcal{K} \to \mathbb{Z}_{\ge 1}$ and assignment map $\pi : \mathcal{K} \times [\sigma(\cdot)] \to [P]$}
$N \leftarrow \sum_{k \in \mathcal{K}} \hat{c}(k)$\;
$L \leftarrow \lceil N/P \rceil$\;
\ForEach{$k \in \mathcal{K}$}{
  $\sigma(k) \leftarrow \lceil \hat{c}(k)/L \rceil$\;
}
$\mathit{splits} \leftarrow \{\langle k,i \rangle : k \in \mathcal{K},\; i \in [0,\sigma(k))\}$\;
\ForEach{$\langle k,i \rangle \in \mathit{splits}$}{
  $s(\langle k,i\rangle) \leftarrow \lceil \hat{c}(k)/\sigma(k) \rceil$\;
}
Sort $\mathit{splits}$ in non-increasing order of $s(\cdot)$; break ties by split label\;
Initialise min-heap $H$ with $P$ empty bins (load\,$=0$, id\,$\in [P]$)\;
\ForEach{$\langle k,i \rangle \in \mathit{splits}$ (sorted)}{
  $b \leftarrow \mathrm{extractMin}(H)$\;
  $\pi(k, i) \leftarrow b.\mathit{id}$\;
  $b.\mathit{load} \leftarrow b.\mathit{load} + s(\langle k,i \rangle)$\;
  $\mathrm{insert}(H, b)$\;
}
\Return $(\sigma, \pi)$\;
\end{algorithm}

\paragraph{Toy example}
Consider four keys $\mathcal{K}=\{A,B,C,D\}$ and $P=2$ workers, with merged frequency estimates $\hat{c}(A)=60$, $\hat{c}(B)=20$, $\hat{c}(C)=12$ and $\hat{c}(D)=8$.
The total is $N=100$, so the target load is $L=\lceil 100/2\rceil=50$.
Applying Equation~\eqref{eq:numsplits}, key $A$ gets $\sigma(A)=\lceil 60/50\rceil=2$ splits, while $B$, $C$ and $D$ each get $\sigma=\lceil \hat{c}/50\rceil=1$ split.
The dominant key $A$ is therefore divided into $\langle A,0\rangle$ and $\langle A,1\rangle$ of size $\lceil 60/2\rceil=30$ each, and the other keys keep a single split of size $20$, $12$ and $8$.

Table~\ref{tab:toy} traces the LPT step.
The five splits are sorted by non-increasing size, and each is assigned to the currently least-loaded worker, with ties broken by worker identifier.
Both workers end with a load of $50$, equal to the mean load $N/P$.
In contrast, any assignment that keeps all events of a key on one worker must place $A$ alone on a worker, giving loads of $60$ and $40$, i.e., a max-to-mean ratio of $1.2$ instead of $1.0$.
At runtime, an event $e$ of key $A$ goes to $\pi(A, \mathit{uid}(e) \bmod 2)$: events with an even identifier go to worker~0 and those with an odd identifier to worker~1, so $A$'s traffic is shared evenly.

\begin{table}[t]
  \centering
  \caption{LPT assignment for the toy example ($P=2$). Loads are shown after each assignment.}
  \label{tab:toy}
  \begin{tabular}{@{}clrcrr@{}}
    \toprule
    Step & Split & Size & Worker & Load $w_0$ & Load $w_1$ \\
    \midrule
    1 & $\langle A,0\rangle$ & 30 & $w_0$ & 30 & 0 \\
    2 & $\langle A,1\rangle$ & 30 & $w_1$ & 30 & 30 \\
    3 & $\langle B,0\rangle$ & 20 & $w_0$ & 50 & 30 \\
    4 & $\langle C,0\rangle$ & 12 & $w_1$ & 50 & 42 \\
    5 & $\langle D,0\rangle$ & 8  & $w_1$ & 50 & 50 \\
    \bottomrule
  \end{tabular}
\end{table}

\paragraph{Resulting partitioning function}
The output of the example is the assignment map $\pi$ together with the split counts $\sigma$:
\begin{align*}
  \sigma &= \{A{:}2,\; B{:}1,\; C{:}1,\; D{:}1\},\\
  \pi &= \{\langle A,0\rangle{\mapsto}0,\; \langle A,1\rangle{\mapsto}1,\; \langle B,0\rangle{\mapsto}0,\\
      &\phantom{={}\{}\langle C,0\rangle{\mapsto}1,\; \langle D,0\rangle{\mapsto}1\}.
\end{align*}
Both are small lookup tables with one entry per key and per split, so $\pi$ has $\sum_k \sigma(k)=5$ entries here and fewer than $|\mathcal{K}|+P$ in general (Section~\ref{sec:analysis}).
Every PartitionerOperator subtask computes them from the same merged counts at each checkpoint (Section~\ref{sec:dynamic}), so all subtasks route identically.
To route an event, a subtask looks up $\sigma(k)$ for its key, computes the split index $i=\mathit{uid}(e)\bmod\sigma(k)$, and returns $\pi(k,i)$:
\begin{itemize}
  \item an event of key $A$ with $\mathit{uid}=7$ has $i=7\bmod 2=1$ and goes to worker~1; with $\mathit{uid}=8$ it has $i=0$ and goes to worker~0;
  \item an event of key $C$ with any $\mathit{uid}$ has $i=\mathit{uid}\bmod 1=0$ and goes to worker~1.
\end{itemize}
Because $\pi$ is a plain map and routing is a lookup plus one modulo, it is cheap to evaluate per event and needs no communication between subtasks.

\subsection{Dynamic Oracle}
\label{sec:dynamic}

\begin{figure}[t]
  \centering
  \begin{tikzpicture}[
      node distance=8mm and 18mm,
      box/.style={draw, rounded corners=3pt, fill=white,
                  minimum height=7mm, minimum width=26mm,
                  font=\scriptsize, align=center},
      arr/.style={-{Stealth[length=4pt]}, thick},
      darr/.style={-{Stealth[length=4pt]}, thick, dashed, gray},
    ]
    \node[box] (src) {Source\\[-1pt]\scriptsize($\times P$)};
    \node[box, below=of src] (col) {CollectorOperator\\[-1pt]\scriptsize($\times P$)};
    \node[box, below=of col] (par) {PartitionerOperator\\[-1pt]\scriptsize($\times P$)};
    \node[box, below=of par] (snk) {Writer\\[-1pt]\scriptsize($\times P$)};

    \draw[arr] (src) -- (col) node[midway, right, font=\scriptsize]{events};
    \draw[arr] (col) -- (par) node[midway, right, font=\scriptsize]{events};
    \draw[arr] (par) -- (snk) node[midway, right, font=\scriptsize]{partitioned events};

    \draw[darr] (col.east) -- ++(14mm,0) |- (par.east)
        node[pos=0.25, right, font=\scriptsize, align=left]{local counts\\(@ checkpoint)}
        node[pos=0.75, above, font=\scriptsize]{broadcast};
  \end{tikzpicture}
  \caption{Flink DAG topology of \textsc{CheckCo}. Each CollectorOperator subtask emits local key counts via a side output at every checkpoint barrier. The PartitionerOperator merges these counts at checkpoint time, recomputes the routing function, and persists the counts in Flink broadcast state.}
  \label{fig:dag}
\end{figure}
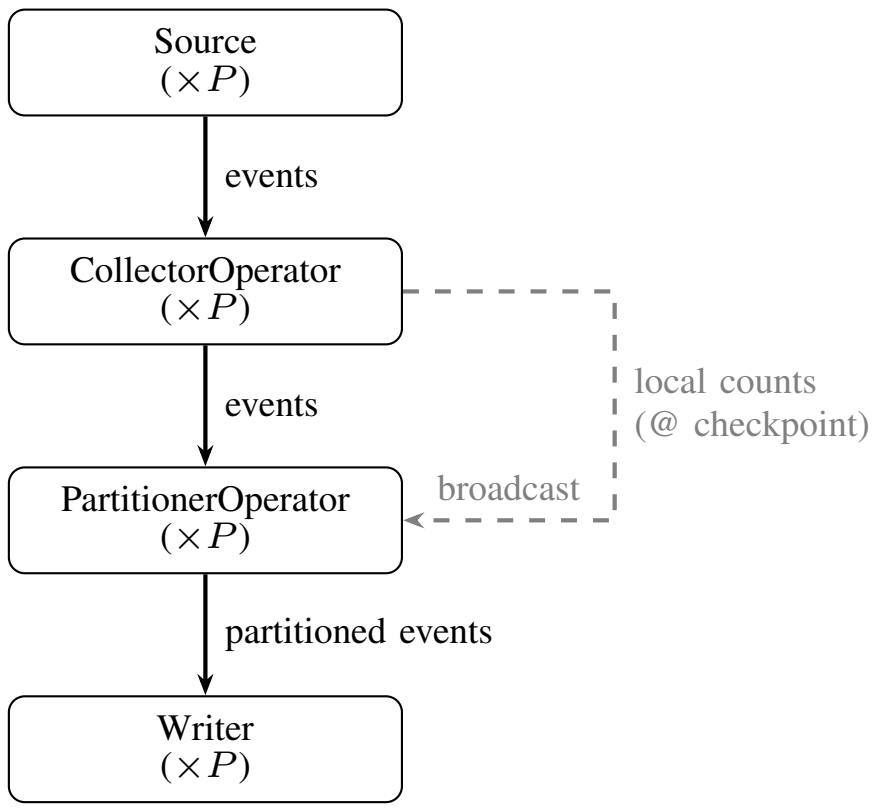

The dynamic oracle builds the frequency map $\hat{c}$ from the live event stream by accumulating per-key counts at each Flink subtask and merging them at every checkpoint.
Figure~\ref{fig:dag} illustrates the Flink DAG topology.

\paragraph{Flink DAG topology}
\textsc{CheckCo} consists of two operators, a \emph{CollectorOperator} and a \emph{PartitionerOperator}, whose handlers are given in Algorithm~\ref{alg:oracle}.
Each CollectorOperator subtask $j$ counts the events of each key in a local map $\lambda_j : \mathcal{K} \to \mathbb{Z}_{\geq 0}$ and passes the events through unchanged; in \texttt{prepareSnapshotPreBarrier}, which runs before the barrier is forwarded, it emits $\lambda_j$ on a \emph{side output} and resets it.
The PartitionerOperator is a \texttt{BroadcastProcessFunction}: \texttt{processElement} routes the events with the current $\pi$, and \texttt{processBroadcastElement} collects the local maps, which are broadcast to all of its subtasks.
In \texttt{snapshotState}, each subtask merges the collected maps, updates the sliding-window buffer of Section~\ref{sec:bootstrap}, runs Algorithm~\ref{alg:core}, and writes the counts to broadcast state.
Every subtask receives the same local maps and runs the same deterministic computation (its only time input is the checkpoint timestamp), so all subtasks compute the same $\pi$ and hold the same oracle state, even though the write happens at snapshot time and not in \texttt{processBroadcastElement}.
A checkpoint aborted after only some subtasks have snapshotted could make the copies diverge; this would affect only the balance, as the writer holds no per-key routing state.

\begin{algorithm}[t]
\caption{Dynamic oracle}
\label{alg:oracle}
\SetKwProg{On}{on}{:}{}
\tcp{CollectorOperator, subtask $j$}
\On{event $e$}{
  $\lambda_j(\mathit{key}(e)) \leftarrow \lambda_j(\mathit{key}(e)) + 1$; forward $e$\;
}
\On{\texttt{prepareSnapshotPreBarrier}}{
  emit $\lambda_j$ on the side output; $\lambda_j \leftarrow \emptyset$\;
}
\tcp{PartitionerOperator, every subtask}
\On{\texttt{initializeState}}{
  restore buckets $M$, counts $\hat{c}$ and EMA slots $\mu$ from broadcast state\;
  \lIf{$\hat{c} \neq \emptyset$}{$(\sigma, \pi) \leftarrow \text{Algorithm~\ref{alg:core}}(\hat{c}, P)$}
  \lElse{$(\sigma, \pi) \leftarrow (\sigma_{\text{cold}}, \pi_{\text{cold}})$ until the first checkpoint}
}
\On{broadcast element $\lambda_j$}{
  append $\lambda_j$ to the list $\Lambda$\;
}
\On{event $e$ of key $k$}{
  \lIf{$\sigma(k)$ is defined}{send $e$ to $\pi(k, \mathit{uid}(e) \bmod \sigma(k))$}
  \lElse{send $e$ to partition $P-1$}
}
\On{\texttt{snapshotState}}{
  $m_1 \leftarrow \sum_{\lambda \in \Lambda} \lambda$; $\Lambda \leftarrow \emptyset$\;
  prepend $m_1$ to $M$, so that $M = \langle m_1, m_2, \ldots \rangle$ is newest first\;
  update $\mu$ with $|m_1|$ events since the last checkpoint\;
  $\theta \leftarrow \bar{\mu} \cdot \Delta_{\mathrm{ckpt}} \cdot \beta$\;
  $B \leftarrow$ the smallest $b \ge B_{\min}$ with $\sum_{i \le b} |m_i| \ge \theta$, or $|M|$ if there is none\;
  $M \leftarrow \langle m_1, \ldots, m_B \rangle$; $\hat{c} \leftarrow \sum_{b=1}^{B} m_b$\;
  $(\sigma, \pi) \leftarrow \text{Algorithm~\ref{alg:core}}(\hat{c}, P)$\;
  write $M$, $\hat{c}$ and $\mu$ to broadcast state under \texttt{"MERGED\_DATA"}\;
}
\end{algorithm}

\paragraph{Checkpoint-synchronised merge}
Because all CollectorOperator subtasks emit their local counters via \texttt{prepareSnapshotPreBarrier} \emph{before} forwarding the checkpoint barrier, and the deployment uses aligned exactly-once checkpoints, the PartitionerOperator is guaranteed to have received all local maps by the time its own \texttt{snapshotState} runs.
The counts of the window are the pointwise sum of all local maps:
\[
  m_1(k) = \sum_{j=1}^{P} \lambda_j(k),
\]
where $\lambda_j$ is the counter map emitted by CollectorOperator subtask $j$; with a single bucket in the buffer of Section~\ref{sec:bootstrap}, $\hat{c} = m_1$.
For a count map $m$, we write $|m| = \sum_k m(k)$ for its total count.

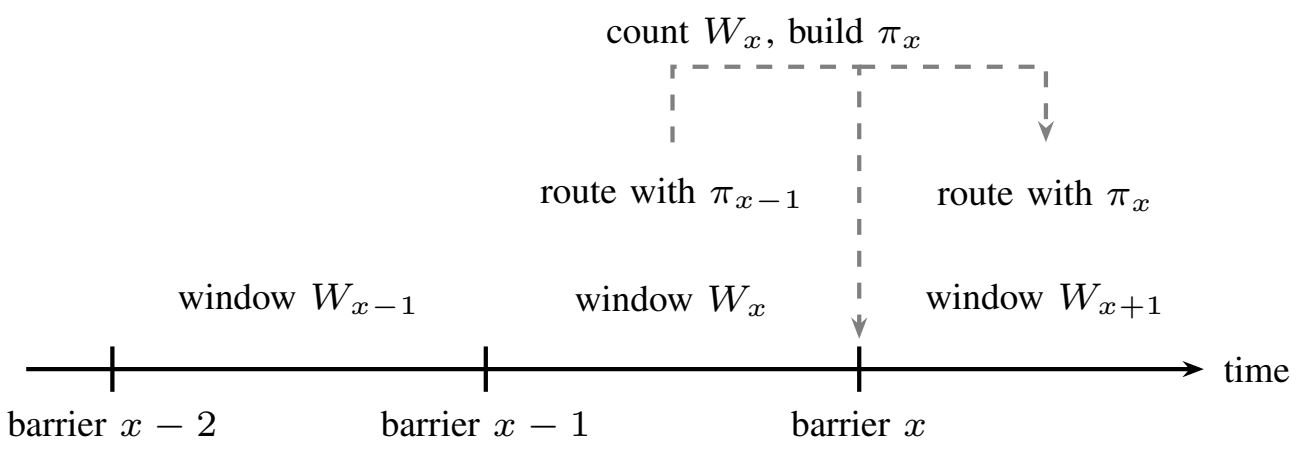
\begin{figure}[t]
  \centering
  \begin{tikzpicture}[
      x=0.95cm, y=1cm,
      bar/.style={thick},
      lab/.style={font=\scriptsize, align=center},
      arr/.style={-{Stealth[length=4pt]}, thick},
    ]
    \draw[arr] (0,0) -- (8.2,0) node[right, font=\scriptsize]{time};
    \foreach \xx/\nm in {0.6/{x-2}, 3.2/{x-1}, 5.8/{x}} {
      \draw[bar] (\xx,-0.15) -- (\xx,0.15);
      \node[lab, below] at (\xx,-0.15) {barrier $\nm$};
    }
    \node[lab] at (1.9,0.45) {window $W_{x-1}$};
    \node[lab] at (4.5,0.45) {window $W_{x}$};
    \node[lab] at (7.1,0.45) {window $W_{x+1}$};
    \node[lab] at (4.5,1.15) {route with $\pi_{x-1}$};
    \node[lab] at (7.1,1.15) {route with $\pi_{x}$};
    \draw[arr, dashed, gray] (4.5,1.5) -- (4.5,2.0) -- (5.8,2.0) -- (5.8,0.2);
    \node[lab, above] at (5.15,2.0) {count $W_x$, build $\pi_x$};
    \draw[arr, dashed, gray] (5.8,2.0) -- (7.1,2.0) -- (7.1,1.5);
  \end{tikzpicture}
  \caption{Timeline of \textsc{CheckCo}. The events of window $W_x$ are counted by key while they are routed with $\pi_{x-1}$. At barrier $x$ the counts are merged and $\pi_x$ is computed, and $\pi_x$ then routes the events of $W_{x+1}$.}
  \label{fig:timeline}
\end{figure}

\paragraph{Windowed operation}
Consecutive checkpoint barriers delimit \emph{checkpoint windows}: window $W_x$ contains the events processed between barrier $x{-}1$ and barrier $x$.
While $W_x$ is processed, every event is routed with the partitioning function $\pi_{x-1}$ that was computed at the previous checkpoint, and the CollectorOperator subtasks count the events of $W_x$ by key.
At barrier $x$, these counts are merged into $\hat{c}_x$, and the PartitionerOperator computes a new partitioning function $\pi_x$ from $\hat{c}_x$ with Algorithm~\ref{alg:core}.
The function $\pi_x$ is then used to route the events of the next window $W_{x+1}$, while the counts of $W_{x+1}$ are collected to build $\pi_{x+1}$ at barrier $x{+}1$.
The partitioning function computed at the end of window $x$ is therefore applied to window $x{+}1$, as illustrated in Figure~\ref{fig:timeline}.
The problem of Section~\ref{sec:problem} is thus solved window by window: each $\pi_x$ approximately minimises the imbalance of the next window under the estimate $\hat{c}_x$.
Since the average load is additive over windows and the maximum of a sum is at most the sum of the maxima, the cumulative imbalance $I(t)$ is at most the sum of the per-window imbalances.
The downstream file writer holds no per-key routing state, so remapping splits between windows is safe.
This cycle repeats at every checkpoint for as long as the job runs, so the routing function follows the traffic of the stream at checkpoint granularity and no offline statistics are needed.
When the sliding-window buffer of Section~\ref{sec:bootstrap} is enabled, $\hat{c}_x$ is computed from the most recent buckets instead of $W_x$ alone, but the timing is unchanged.
The first window $W_1$ has no previous function and is routed by the cold-start or bootstrap mechanisms of Section~\ref{sec:bootstrap}.
Because $\pi_x$ is built from the traffic of $W_x$ and applied to $W_{x+1}$, \textsc{CheckCo} assumes that the key distribution changes little from one window to the next; a shorter checkpoint interval makes this assumption easier to satisfy, at the cost of more frequent merges.

\paragraph{State and recovery}
The oracle state, i.e., the buffer of recent buckets, the merged counts $\hat{c}$ and the EMA slots of Section~\ref{sec:bootstrap}, is stored in Flink broadcast state under a fixed key \texttt{"MERGED\_DATA"}; $\pi$ itself is not stored.
On restart after a failure, each PartitionerOperator subtask restores this entry and recomputes $\pi$ from $\hat{c}$ with Algorithm~\ref{alg:core}; because the algorithm is deterministic, this yields the function in force at the last completed checkpoint, without contacting any external service.
Broadcast state is held in memory at runtime, and the deployment uses a heap-based state backend; the oracle state holds one count per key and bucket plus $1{,}440$ EMA slots, and its size is independent of the event rate.

\subsection{Bootstrapping}
\label{sec:bootstrap}

The oracle passes through three distinct phases depending on when and how it is initialised.

\paragraph{Cold-start (no prior statistics)}
Before the first checkpoint completes, no oracle estimate is available.
In this period, events are routed by their key identifier (the integer id of the key) modulo $P$, that is, every key has a single split:
\[
  \sigma_{\text{cold}}(k) = 1, \qquad \pi_{\text{cold}}(k, 0) = k \bmod P.
\]
This spreads keys uniformly across the $P$ partitions in expectation, but assigns each key to exactly one partition with no splitting.
The cold-start period therefore exhibits the same per-key concentration as \textsc{KeyBy}: if the key distribution is skewed, so is the load during this phase.
Oracle-driven split-based routing takes effect after the first checkpoint completes, at which point the partitioning function is recomputed and imbalance is corrected.

\paragraph{Static bootstrap (from historical statistics)}
The cold-start phase can be eliminated by pre-loading $\hat{c}$ from an external source before the pipeline starts.
A natural choice is the key frequency counts of prior runs, kept in any external service or storage; the operator reads them at startup and writes the resulting frequency map directly into the PartitionerOperator's broadcast state as the initial $\hat{c}$.
The pipeline then starts with a well-calibrated partitioning function and incurs no imbalanced cold-start period.
The dynamic oracle continues to update $\hat{c}$ at every subsequent checkpoint, so the static bootstrap merely seeds the oracle; it does not replace the adaptive mechanism.

\paragraph{Warm restart (after failure)}
When a Flink job restarts after a failure, all operator state is restored from the most recent successful checkpoint.
The PartitionerOperator restores the counts from broadcast state (keyed by \texttt{"MERGED\_DATA"}), recomputes $\pi$, and resumes routing immediately with the pre-failure partitioning function.
No cold-start phase occurs and no external system is queried; recovery is bounded by Flink's checkpoint restore time.

\paragraph{Sliding-window buffer}
A single checkpoint's worth of counts may be statistically noisy if the checkpoint interval is short or traffic is low.
We maintain a buffer of the $B$ most recent checkpoint buckets and compute the oracle estimate from their union:
\[
  \hat{c}(k) = \sum_{b=1}^{B} m_b(k),
\]
where $m_b(k)$ is the count for key $k$ in bucket $b$; this sum is the estimate $\hat{c}_x$ used at checkpoint $x$ in the windowed operation above.
At each checkpoint, the newest bucket is added, and the buffer keeps the shortest run of most recent buckets, at least $B_{\min}$ of them, whose total count $\sum_k \hat{c}(k)$ meets a minimum threshold $\theta$; older buckets are discarded (Algorithm~\ref{alg:oracle}).

\paragraph{Adaptive minimum buffer size}
The threshold $\theta$ is calibrated to current traffic throughput using a per-minute moving average.
The system maintains a 1440-entry circular buffer (one entry per minute of the day).
Each entry stores an exponential moving average (EMA) of the per-minute event rate observed at that minute-of-day:
\[
  \mu_t \leftarrow \alpha \cdot v_t + (1 - \alpha) \cdot \mu_t,
\]
where $v_t$ is the per-minute event rate observed at the current checkpoint (the size of the newest bucket divided by the minutes since the previous checkpoint), applied to every minute slot covered since that checkpoint, $\alpha \in (0,1)$ is a smoothing factor, and $\mu_t$ is the prior estimate for that minute slot.
The minimum buffer size is then:
\[
  \theta = \bar{\mu} \cdot \Delta_{\mathrm{ckpt}} \cdot \beta,
\]
where $\bar{\mu}$ is the average of a symmetric window of minute slots centred on the current time-of-day (a window of $360$ minutes, i.e., $\pm 180$ minutes, in our deployment), $\Delta_{\mathrm{ckpt}}$ is the configured checkpoint interval in minutes, and $\beta$ is a scaling factor.
Until all 1440 slots have been populated, which takes about a day, $\bar{\mu}$ is instead the average of the populated slots over the whole day.
This mechanism tracks daily traffic patterns: during peak hours, each bucket contains more events so fewer buckets suffice; during off-peak hours, more buckets are retained to maintain statistical stability.

%% file: sections/analysis.tex
\section{Analysis}
\label{sec:analysis}

\paragraph{Hardness}
Minimising the imbalance when every key is kept on a single worker is the scheduling problem $P\|C_{\max}$, which is strongly NP-hard when $P$ is part of the input~\cite{Garey:1978:Strong}.
Shuffle grouping balances the load trivially, but at fan-out $P$ for every key.
\textsc{CheckCo} packs splits of fixed size, which is again $P\|C_{\max}$, now over the splits; it therefore uses LPT, and the bounds below are approximation guarantees on the load and on the fan-out of the function it constructs.

\paragraph{2-approximation relative to the optimum with freely split keys}
All bounds in this section are stated with respect to the estimated counts $\hat{c}$, not the true loads; the $\hat{c}(k)$ are non-negative integers, and we assume $N \ge 1$.
Let $C_{\max}^{*}$ denote the makespan of the optimal assignment of the $N$ estimated events to $P$ partitions, allowing any key to be split arbitrarily; some partition receives at least $\lceil N/P \rceil$ events, so $C_{\max}^{*} \ge L$.
Let $C_{\max}^{\mathrm{LPT}}$ denote the makespan of Algorithm~\ref{alg:core} measured in the split sizes $s(\cdot)$ that LPT packs.
Graham's list-scheduling bound~\cite{Graham:1966:Bounds} states that any greedy assignment that places each job on the currently least loaded machine, LPT included, has makespan at most the average load plus the largest job.
Every split has size at most $L$; keys with $\hat{c}(k) = 0$ get no splits, and for every key with $\hat{c}(k) > 0$, since $\sigma(k) < \hat{c}(k)/L + 1$, the ceiling adds at most $\sigma(k) - 1 < \hat{c}(k)/L$ to the packed size of key $k$, so the packed sizes sum to less than $N + N/L \le N + P$.
Hence $C_{\max}^{\mathrm{LPT}} < (N+P)/P + L \le 2L + 1$; as the makespan is an integer, $C_{\max}^{\mathrm{LPT}} \le 2L \le 2\,C_{\max}^{*}$.
If each split receives at most $s(\cdot)$ events (see hash uniformity below), the realised load of every partition is at most its packed load, so the bound also holds for the realised makespan.

\paragraph{Tighter 4/3-approximation relative to the split optimum}
Let $C_{\max}^{\mathrm{split}}$ denote the optimal makespan for the set of splits produced by Algorithm~\ref{alg:core} (i.e., treating splits as fixed jobs).
Graham's LPT algorithm~\cite{Graham:1969:LPT} guarantees:
\[
  C_{\max}^{\mathrm{LPT}} \leq \left(\frac{4}{3} - \frac{1}{3P}\right) C_{\max}^{\mathrm{split}}.
\]
This bound is relative to the optimum of the generated splits and is not directly comparable with the 2-approximation above: it shows that LPT is near-optimal for the splits it is given, while the split construction (Equation~\eqref{eq:numsplits}) keeps $C_{\max}^{\mathrm{split}} \le C_{\max}^{\mathrm{LPT}} \le 2\,C_{\max}^{*}$ by bounding every split size by $L$.

\paragraph{Fan-out}
Splitting is the price of balance (Section~\ref{sec:problem}); we bound it.
Consider the keys with $\hat{c}(k) > 0$.
A key occupies at most as many partitions as it has splits, so $\varphi_\pi(k) \le \sigma(k) < \hat{c}(k)/L + 1$.
Summing over keys and using $L \ge N/P$ gives
\[
  \sum_{k} \varphi_\pi(k) < |\mathcal{K}| + N/L \le |\mathcal{K}| + P.
\]
Any partitioning function sends every such key to at least one worker, so its total fan-out is at least $|\mathcal{K}|$.
\textsc{CheckCo} is therefore within an additive $P$ of the minimum, and within a factor $2$ when $|\mathcal{K}| \ge P$.
As with the other bounds, this holds on estimated counts.

\paragraph{Hash uniformity}
The split index $i = \mathit{uid}(e) \bmod \sigma(k)$ assumes that, within each key, the residues of $\mathit{uid}(e)$ modulo $\sigma(k)$ are close to uniform.
If the $c$ events of key $k$ routed in window $W_{x+1}$ have consecutive identifiers, every split receives $\lfloor c/\sigma(k) \rfloor$ or $\lceil c/\sigma(k) \rceil$ of them, and with uniform residues this holds in expectation; $c$ differs from $\hat{c}_x$ by the estimation error, not by rounding.
The per-subtask loads of Section~\ref{sec:eval} ($112$ of $128$ subtasks within $-3.1\%$ to $+1.4\%$ of the mean) are consistent with this assumption.

%% file: sections/lakehouse.tex
\section{Application: Ingestion into a Data Lakehouse}
\label{sec:lakehouse}

One application of \textsc{CheckCo} is streaming ingestion into a data lakehouse.
In such a pipeline, the event stream is written as serialised files to cloud object storage, where they are registered in a data lakehouse table, in an open table format such as Apache Iceberg~\cite{Iceberg:Spec}, that query engines such as Snowflake~\cite{Dageville:2016:Snowflake} read by key and time range.
Each writer subtask rolls its files at every checkpoint and writes one file per key it received, so the number of files for a key in a checkpoint interval equals the number of subtasks that receive it.
The routing function therefore also decides how many files a query must open.

Under key-based routing, each key produces one file per checkpoint, but a heavy key produces one very large file from a single overloaded subtask.
Under shuffle routing, each key can reach all $P$ subtasks, giving up to $P$ files per key and $|\mathcal{K}|\cdot P$ files per checkpoint, most of them small.
Under split-based routing, key $k$ produces at most $\sigma(k)$ files, i.e., $\varphi_\pi(k) \le \sigma(k)$ in the notation of Section~\ref{sec:problem}, so the fan-out bound of Section~\ref{sec:analysis} also bounds the number of files.
A key with $\hat{c}(k) \le L$ has $\sigma(k)=1$ and a single file, and only the fewer than $N/L \le P$ heavy keys are split, into at most $P$ files of similar size.
A checkpoint thus produces at most $\sum_k \sigma(k) < |\mathcal{K}| + P$ files, fewer than $P$ more than the key-based minimum and far fewer than shuffling.
For a query on one key, this means one file per interval for light keys and a handful of equal-sized files, which can be scanned in parallel, for heavy keys.

%% file: sections/experiments.tex
\section{Experimental Evaluation}
\label{sec:eval}

\subsection{Application}
\label{sec:app}

\begin{figure*}[t]
  \centering
  \begin{tikzpicture}[
      node distance=4.5mm,
      box/.style={draw, rounded corners=3pt, fill=white, minimum height=9mm,
                  text width=17mm, inner xsep=2pt, font=\scriptsize, align=center},
      store/.style={box, rounded corners=0pt, fill=gray!10},
      arr/.style={-{Stealth[length=4pt]}, thick},
      darr/.style={-{Stealth[length=4pt]}, thick, dashed, gray},
    ]
    \node[store] (mq) {Message queue\\(512 partitions)};
    \node[box, right=of mq] (src) {Source\\($\times P$)};
    \node[box, right=of src] (val) {Validation and\\type filter ($\times P$)};
    \node[box, right=of val, fill=green!10, text width=20mm] (col) {CollectorOperator\\($\times P$)};
    \node[box, right=of col, fill=green!10, text width=20mm] (par) {PartitionerOperator\\($\times P$)};
    \node[box, right=9mm of par] (wr) {File writer\\($\times P$)};
    \node[store, right=of wr] (os) {Cloud object\\storage};
    \node[box, below=5mm of val, minimum height=6mm, dashed] (inv) {Invalid events};
    \coordinate (bc) at ($(col.south)!0.5!(par.south)+(0,-4mm)$);
    \node[font=\scriptsize, gray, below=0.5mm of bc] (bcl) {local counts, broadcast at each checkpoint};
    \begin{scope}[on background layer]
      \node[draw, dashed, rounded corners=4pt, inner sep=2mm, fit=(col)(par)(bcl),
            label={[font=\scriptsize]above:\textsc{CheckCo} (Section~\ref{sec:dynamic})}] {};
    \end{scope}
    \draw[arr] (mq) -- (src);
    \draw[arr] (src) -- (val);
    \draw[arr] (val) -- (col);
    \draw[arr] (col) -- (par);
    \draw[arr] (par) -- node[above, font=\scriptsize]{$\pi$} (wr);
    \draw[arr] (wr) -- (os);
    \draw[arr, dashed] (val) -- (inv);
    \draw[darr] (col.south) |- (bc) -| (par.south);
  \end{tikzpicture}
  \caption{The event ingestion application. \textsc{CheckCo} routes the valid events to the subtasks of the file writer; the routing function $\pi$ is recomputed at every checkpoint.}
  \label{fig:app}
\end{figure*}
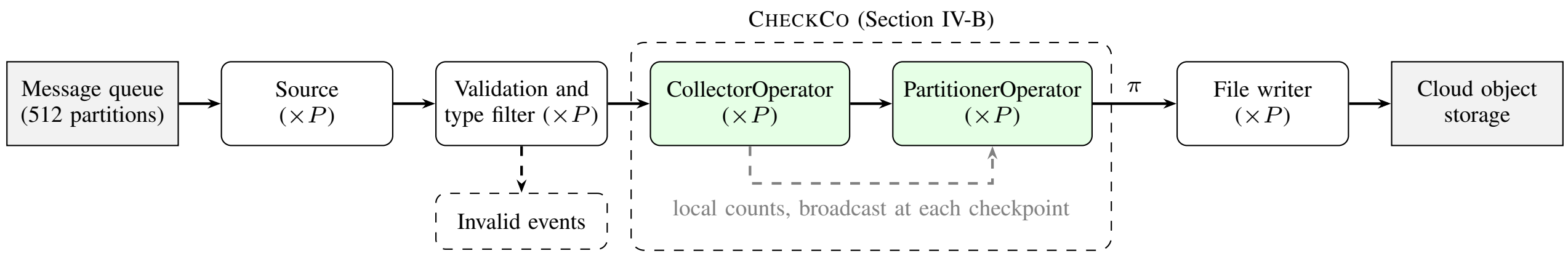

We evaluate \textsc{CheckCo} on the event ingestion pipeline of one of our largest games, which moves its telemetry events from a message queue into cloud object storage (Figure~\ref{fig:app}).
Each event carries its event type, which is the key, a timestamp and a payload.
A Flink job reads the events from the message queue, parses and validates them, and drops the events of blocklisted types; events that fail validation go to a side output and are stored separately.
The CollectorOperator and PartitionerOperator of Section~\ref{sec:dynamic} then route the valid events to the $P$ subtasks of the file writer, which serialises them into files in cloud object storage and commits the files when a checkpoint completes.
Most of the work of the job is done by the file writer, so its subtasks are the workers whose loads $\ell_p(t)$ \textsc{CheckCo} balances.

\subsection{Setup}

All experiments were conducted on a production event ingestion Flink cluster managed by the Flink Kubernetes Operator.
The cluster comprises 16 TaskManagers with 8 task slots each (parallelism $P = 128$), running Apache Flink 2.1.3.
Each TaskManager pod is allocated 16\,GB of memory and 8 vCPUs;
the JobManager pod uses 8\,GB and 4 vCPUs.
The event source is a message queue with 512 partitions.

\paragraph{Data}
We report measurements from two production days of this stream.
On 6 October 2026 (UTC), the job ran for the whole day without a restart; we use this day for the steady-state evaluation.
On 17 October 2026, the job was restarted three times, which lets us observe \textsc{CheckCo} during recovery.
We report the load of each file writer subtask, as the number of events it receives, and the daily number of events of each key.
On 6 October, the stream carried over 100 billion events, with a peak rate of $1.75 \times 10^6$ events/s.

\paragraph{Baseline}
\textsc{KeyBy} uses Flink's built-in \texttt{keyBy} operator to route all events of the same key to the same subtask.
On the observed per-key counts, this would create severe imbalance: subtasks handling dominant keys would process several times the mean load (estimated by simulation below).

\paragraph{Algorithm under test}
\textsc{CheckCo} is configured with checkpoint interval $\Delta_{\mathrm{ckpt}} = 10$ minutes,
parallelism $P = 128$, minimum bucket count $B_{\min} = 1$,
EMA smoothing factor $\alpha = 0.25$, and buffer-size scaling factor $\beta = 0.5$.
The EMA operates on a 1{,}440-slot circular buffer (one slot per minute of the day),
with a 360-minute averaging window centred on the current time of day ($\pm 180$ minutes) for computing the minimum buffer size threshold.

\paragraph{Metrics}
To compare days and intervals of different volume, we report the imbalance $I(t)$ of Section~\ref{sec:problem} relative to the average load, as the \emph{imbalance ratio}
\[
  R(t) = \frac{\max_p \ell_p(t)}{\operatorname*{avg}_p \ell_p(t)} = 1 + \frac{I(t)}{\operatorname*{avg}_p \ell_p(t)},
\]
where $R(t) = 1$ is perfect balance; since the average load does not depend on the partitioning function, minimising $R(t)$ is the same problem as minimising $I(t)$.
\begin{itemize}
  \item \emph{Cumulative load imbalance ratio}: $R(t)$, where $\ell_p(t)$ is the number of events received by subtask $p$ since 00:00 UTC.
  \item \emph{Per-interval load imbalance ratio}: the same ratio computed over the events of each 5-minute interval. It shows how the balance evolves within the day.
  \item \emph{Throughput}: events processed per second, aggregated across all subtasks.
\end{itemize}
Both imbalance ratios are measured on actual loads, not on the oracle estimates.
We do not report per-event latency: the file writer commits its files when a checkpoint completes, so the time until an event becomes visible in object storage is governed by the checkpoint interval rather than by the routing.

\subsection{Skew Characterisation}

\begin{figure}[t]
  \centering
  \includegraphics[width=\columnwidth]{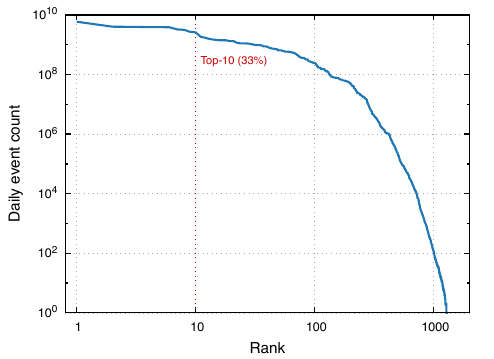}
  \caption{Per-key event counts on 6 October 2026 (log-log scale, $|\mathcal{K}| = 1{,}315$ keys).
           The top-10 keys account for $32.9\%$ of all events.}
  \label{fig:skew}
\end{figure}

Figure~\ref{fig:skew} shows the rank-frequency distribution of per-key event counts on 6 October.
The stream contains $|\mathcal{K}| = 1{,}315$ keys (event types), and the distribution is heavy-tailed.
The top-10 keys account for $32.9\%$ of all events, and the most frequent key alone for $5.2\%$, about $69\times$ the per-key mean and $2.6 \times 10^5\times$ the median key.
On the daily counts, only $36$ keys exceed the target load $L = \lceil N/P \rceil$; these are the keys that \textsc{CheckCo} splits.

This skew directly motivates split-based partitioning.
Under \textsc{KeyBy}, the subtask that receives the most frequent key would carry at least $6.7\times$ the mean subtask load from that key alone (computed from the per-key counts).
\textsc{CheckCo} instead gives heavy keys multiple splits, roughly in proportion to their estimated count (Equation~\eqref{eq:numsplits}).

\subsection{\textsc{CheckCo} vs.\ \textsc{KeyBy}}
\label{sec:keyby}

\begin{figure}[t]
  \centering
  \includegraphics[width=\columnwidth]{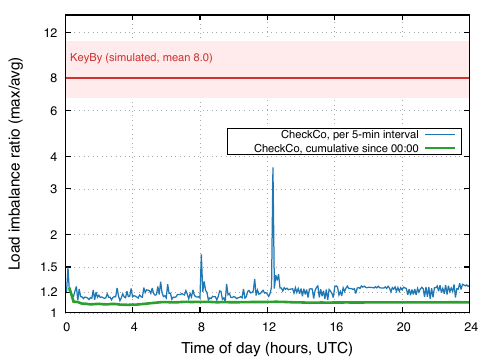}
  \caption{Load imbalance ratio on 6 October 2026 (log scale).
           Blue: \textsc{CheckCo}, measured per 5-minute interval.
           Green: \textsc{CheckCo}, measured on the cumulative load since 00:00 UTC.
           Red: \textsc{KeyBy}, simulated on the same day's per-key daily totals, i.e., a daily (cumulative) ratio (line = mean, band = p5--p95 over trials).}
  \label{fig:imbalance}
\end{figure}

\begin{figure}[t]
  \centering
  \includegraphics[width=\columnwidth]{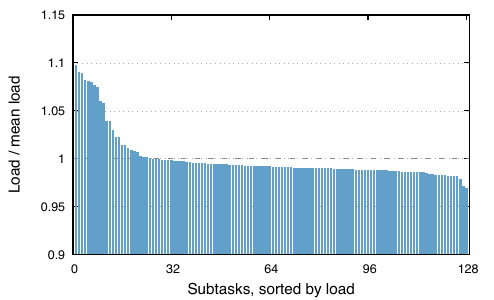}
  \caption{Cumulative load of each of the $P=128$ subtasks over 6 October 2026, divided by the mean and sorted in decreasing order.}
  \label{fig:subtask}
\end{figure}

Figure~\ref{fig:imbalance} shows both imbalance ratios of \textsc{CheckCo} on 6 October, alongside the simulated \textsc{KeyBy} range (see below).
The cumulative ratio, which measures the imbalance minimised in Section~\ref{sec:problem}, settles within the first hour and stays between $1.07$ and $1.10$ for the rest of the day, ending at $1.10$.
This day is typical: over the 30 days from 18 September to 17 October 2026, the daily cumulative ratio has a median of $1.09$ and ranges from $1.04$ to $1.12$, and the job was restarted $85$ times on $26$ of these days without a visible effect on the ratio (median $1.09$ both on days with restarts and on the four days without).
The per-interval ratio has a mean of $1.21$ and a p99 of $1.42$; $97\%$ of the intervals are at or below $1.3$.
Figure~\ref{fig:subtask} shows the end-of-day load of every subtask: $112$ of the $128$ subtasks are within $-3.1\%$ and $+1.4\%$ of the mean, and the most loaded subtask carries $9.8\%$ more than the mean.
The $16$ most loaded subtasks on this day are those with the highest indices ($112$--$127$); the simulation of Section~\ref{sec:baselines} also places its most loaded subtasks at the highest indices from the per-key counts alone ($13$ of its $16$ most loaded are among $112$--$127$), which indicates that the pattern stems from the deterministic packing of Algorithm~\ref{alg:core} rather than from the production environment.

\begin{table}[t]
  \caption{Comparison of \textsc{KeyBy} and \textsc{CheckCo} on 6 October 2026.
           \textsc{CheckCo} values are measured in production; \textsc{KeyBy} imbalance is a simulation estimate (see text);
           \textsc{KeyBy} throughput was not measured.}
  \label{tab:comparison}
  \begin{tabular}{@{}p{3.5cm}cc@{}}
    \toprule
    Metric & \textsc{KeyBy} & \textsc{CheckCo} \\
    \midrule
    Imbalance, cumulative 24\,h & $8.03$ (sim.)  & $1.10$ \\
    \quad p99 over sim.\ trials & $12.67$ (sim.) & --- \\
    Imbalance, 5-min (mean)      & ---  & $1.21$ \\
    Imbalance, 5-min (p99)       & --- & $1.42$ \\
    Throughput, peak (events/s) & ---   & $1.75 \times 10^6$ \\
    \bottomrule
  \end{tabular}
\end{table}

\paragraph{KeyBy baseline}
Because \textsc{CheckCo} has been running in production for over a year, no \textsc{KeyBy} trace of this stream is available for direct measurement.
We therefore estimate the \textsc{KeyBy} imbalance by simulation on the per-key daily totals of the same day, which gives a ratio comparable to \textsc{CheckCo}'s cumulative ratio over the day:
we assign each key to one of $P=128$ subtasks uniformly at random (modelling Flink's hash-based \texttt{keyBy}),
accumulate per-subtask loads, and compute the imbalance ratio.
Over 3{,}000 independent trials this yields a mean imbalance ratio of $8.03$ (p99 over trials: $12.67$), against $1.10$ for \textsc{CheckCo},
with a floor of $6.7\times$ imposed by the most frequent key alone.
At this ratio, the most loaded subtask would have to process about eight times the mean load to keep up with the stream.

\subsection{Comparison with Other Partitioners}
\label{sec:baselines}

We also compare \textsc{CheckCo} with the partitioners of our earlier work by simulation on the same day's per-key distribution.
We draw a stream of $10^7$ events independently from the per-key daily totals of 6 October and route it with each partitioner to $P=128$ workers, over three random seeds.
PKG~\cite{Nasir:2015:PBC:ICDE} sends each event to the less loaded of two hashed workers.
The ICDE 2016 algorithms~\cite{Nasir:2016:WTC:ICDE} treat the keys with frequency at least $\theta = 1/(5P)$ (their default) as the head and the rest as the tail and route the tail with PKG: D-Choices sends each head event to the least loaded of $d$ hashed workers, with a single $d$ for the whole head computed by the procedure of that paper ($d=7$ here), and W-Choices to the least loaded of all workers.
The load-aware partitioners see the exact global worker loads, which favours them.
\textsc{CheckCo} runs Algorithm~\ref{alg:core} on the expected key counts.
Table~\ref{tab:baselines} reports the imbalance ratio at the end of the stream and the total fan-out $\sum_k \varphi_\pi(k)$ over all $1{,}315$ keys of the day.
The fan-out is computed from the partitioning function and not from the sampled stream: a sample of $10^7$ events contains only about $730$ of the keys, since $606$ keys have fewer daily events than one in $10^7$ of the stream, and these keys together carry less than $10^{-5}$ of the events, so leaving them out of the sample does not change the loads.

\begin{table}[t]
  \centering
  \caption{Simulated partitioners on the per-key distribution of 6 October 2026 ($P=128$, $10^7$ events, mean over three seeds). All partitioners use the true key frequencies of the day: \textsc{CheckCo} the exact counts, D-Choices and W-Choices the exact head keys and global worker loads. Fan-out is the number of workers to which the partitioning function may send a key, summed over all 1{,}315 keys of the day, so the \textsc{KeyBy} fan-out is the minimum; \textsc{KeyBy} and \textsc{CheckCo} send events to all of these workers, the load-aware partitioners may use fewer. $^{*}$Mean over $3{,}000$ random assignments of keys to workers (Section~\ref{sec:keyby}) instead of three seeds, because the \textsc{KeyBy} imbalance depends only on how the keys are assigned.}
  \label{tab:baselines}
  \begin{tabular}{@{}lrr@{}}
    \toprule
    Partitioner & Imbalance ratio & Total fan-out \\
    \midrule
    \textsc{KeyBy}                     & $8.03^{*}$  & $1{,}315$ \\
    PKG~\cite{Nasir:2015:PBC:ICDE}      & $3.33$  & $2{,}619$ \\
    D-Choices~\cite{Nasir:2016:WTC:ICDE} & $1.01$ & $3{,}136$ \\
    W-Choices~\cite{Nasir:2016:WTC:ICDE} & $1.00$ & $15{,}977$ \\
    Shuffle                             & $1.00$  & $168{,}320$ \\
    \textsc{CheckCo}, $L = N/P$             & $1.10$  & $1{,}378$ \\
    \textsc{CheckCo}, $L = N/(2P)$        & $1.11$  & $1{,}491$ \\
    \textsc{CheckCo}, $L = N/(4P)$        & $1.03$  & $1{,}712$ \\
    \textsc{CheckCo}, $L = N/(8P)$        & $1.02$  & $2{,}202$ \\
    \bottomrule
  \end{tabular}
\end{table}

PKG cannot balance this stream: the most frequent key carries $5.2\%$ of the events, more than the $2/P \approx 1.6\%$ that two workers can absorb, and its imbalance ratio of $3.33$ matches this bound.
W-Choices reaches near-perfect balance by spreading each of the $106$ head keys over all workers, which multiplies the fan-out by about $12$ with respect to \textsc{CheckCo}; shuffle grouping does the same for every key.
D-Choices balances the stream ($1.01$) with $7$ workers for every head key, at $2.3\times$ the fan-out of \textsc{CheckCo}, although $70$ of the $106$ head keys fit within the target load and keep a single split in \textsc{CheckCo}.
In this simulation, \textsc{CheckCo} keeps the imbalance ratio at $1.10$ with a fan-out within $5\%$ of the \textsc{KeyBy} minimum, because only the keys above the target load $L$ are split, each into roughly as many splits as its load requires.
This simulated ratio equals the one measured in production, which suggests that its remaining imbalance comes mainly from the granularity of the splits rather than from the lag between estimation and use.
A finer target load reduces this remaining imbalance at a moderate cost in fan-out: with $L = N/(4P)$ the ratio drops to $1.03$ at $1.24\times$ the fan-out, and with $L = N/(8P)$ to $1.02$ at $1.6\times$, a fan-out about $30\%$ below that of D-Choices and about $7\times$ below that of W-Choices.
The effect is not monotonic, as $L = N/(2P)$ is slightly worse than $L = N/P$: the remaining imbalance depends on how the split sizes fit together in the packing, and with $L = N/(2P)$ the most loaded worker receives one split of about half the mean load and two of about a third.
This comparison is a one-shot simulation on a stationary stream: \textsc{CheckCo} receives the exact key counts once, as a static oracle would, so it shows only the balance and fan-out of the routing function itself.
It does not evaluate the dynamic oracle, and no trace of a static oracle on this stream is available to compare against in production.
In production, the dynamic oracle removes the need for prior statistics: the counts are built from the live stream, carried across restarts in broadcast state, and over 30 days kept the daily ratio between $1.04$ and $1.12$ (Section~\ref{sec:keyby}).

\subsection{Adapting to Changes in the Key Distribution}
\label{sec:drift}

Because $\pi_x$ is built from $W_x$ and applied to $W_{x+1}$ (Section~\ref{sec:dynamic}), a shift in the key distribution within a window is reflected in routing from the next checkpoint on.
On 6 October, the largest per-interval ratio, $3.63$ in 12:15--12:20 UTC, lasted one interval; the ratio stayed between $1.24$ and $1.41$ until the 12:42 checkpoint and was then back at its usual level.
Such episodes last at most a few checkpoint intervals and barely affect the cumulative ratio, which stays at $1.10$ across this one; a shorter checkpoint interval would shorten them further, at the cost of more frequent merges.

\subsection{Overhead}

On 6 October, all $143$ checkpoints completed, none failed, with a median duration of $6.2$\,s (p99: $12.8$\,s) and a median total checkpoint size of $6.9$\,MB, which includes the oracle state.
Over the day, the PartitionerOperator subtasks were busy $3.0\%$ of the time on average, and the source chain that includes the CollectorOperator $4.4\%$, against $15.6\%$ for the file writer.
On the daily counts, the routing table has $1{,}379$ entries: $1{,}315$ keys, of which $36$ are split, into at most $7$ splits; LPT places two splits of one key on the same worker, so the fan-out is $1{,}378$ (Table~\ref{tab:baselines}).
Because the oracle and routing state grow with the number of keys, buckets and splits and not with the event rate, they stay small at $1.75 \times 10^6$ events/s.

\subsection{Recovery after Restarts}

\begin{figure}[t]
  \centering
  \includegraphics[width=\columnwidth]{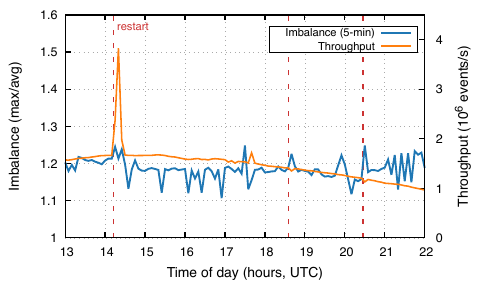}
  \caption{Per-interval imbalance and throughput on 17 October 2026; red dashed lines mark job restarts.}
  \label{fig:restart}
\end{figure}

On 17 October, the job was restarted at 14:12, 18:35 and 20:27 UTC.
After each restart, the partitioning function is recomputed from the counts restored from broadcast state (warm restart, Section~\ref{sec:bootstrap}), so routing resumes with the pre-failure function instead of cold-start routing.
Figure~\ref{fig:restart} shows the imbalance and throughput around the restarts.
For comparison, the cold-start rule ($\mathit{key}(e) \bmod P$), computed on the per-key counts of 6 October, would give a ratio of $8.15$; no interval near that level appears after any of the restarts.
After the first restart, the pipeline processes its backlog at up to $3.83 \times 10^6$ events/s, more than twice the normal rate, while the per-interval imbalance stays between $1.13$ and $1.25$.
Over the whole day, the cumulative ratio is $1.07$ and the per-interval ratio has a mean of $1.23$ (p99: $1.34$), in line with the restart-free day.

%% file: sections/related.tex
\section{Related Work}
\label{sec:related}

\paragraph{Power-of-choices partitioning}
PKG~\cite{Nasir:2015:PBC:ICDE} introduced power-of-$d$-choices ($d=2$) for stream partitioning: each event is routed to the less loaded of $d$ candidate workers chosen by hashing its key.
For stateful operators, it requires the partial per-key results of the two workers to be merged downstream.
Because each key is spread over at most two workers, PKG cannot balance a key that carries more than $2/P$ of the stream.
Our ICDE 2016 algorithms~\cite{Nasir:2016:WTC:ICDE}, D-Choices and W-Choices, extend PKG: they track the most frequent keys (the head) online with Space Saving and route each head event to the least loaded of $d$ hashed workers in D-Choices, with a single $d$ for the whole head, or of all workers in W-Choices; the other keys keep two choices.
FlexD~\cite{Liu:2024:FlexD} follows the same line: it applies key grouping to infrequent keys and dynamically expands the set of instances serving high-frequency keys, trading load balance against aggregation cost, and is evaluated on Apache Storm.
\textsc{CheckCo} addresses the same problem with a dynamic oracle that estimates the frequency of every key, not only the head, merges the local counts of all subtasks at each checkpoint and recovers them from Flink state, and targets a different production setting (event-ingestion file writers with no keyed state to migrate).

In terms of key fan-out, these methods give heavy keys either too few workers (two in PKG) or the same number of workers regardless of their load (a single $d$ for all head keys in D-Choices, all workers in W-Choices), whereas \textsc{CheckCo} chooses the number of splits of each key from its estimated count, so each key receives roughly as many workers as its load requires.
In simulation, this keeps the fan-out of \textsc{CheckCo} within $5\%$ of the \textsc{KeyBy} minimum, less than half that of D-Choices and an order of magnitude below W-Choices (Section~\ref{sec:baselines}).

\paragraph{Balls-into-bins foundations}
Azar et al.~\cite{Azar:1999:BA} showed that placing each ball into the least loaded of $d=2$ randomly chosen bins reduces the maximum load from $\Theta(\log n / \log \log n)$ to $\Theta(\log \log n)$ with high probability, an exponential improvement over a single random choice, and Mitzenmacher~\cite{Mitzenmacher:2001:PTW} extended the analysis to dynamic settings.
PKG adapts this result to stream partitioning; \textsc{CheckCo} instead computes a deterministic split-assignment at checkpoint time, so runtime routing requires only a table lookup and a modulo operation, with no per-event random draw.

\paragraph{Stateful stream partitioning}
Gedik~\cite{Gedik:2014} studies partitioning functions for stateful data parallelism with large, skewed key domains.
His functions are hybrids of an explicit routing table for the most frequent keys and consistent hashing for the rest; frequencies are tracked online with lossy counters over a window, and the adaptation algorithms trade load balance against the state that must migrate when the function changes.
Because the state of a key lives on one worker, every key is mapped to a single partition, so a key that carries more than $1/P$ of the stream cannot be balanced.
Our scheme targets downstream operators without keyed state (file writers), so there is no state to migrate: the partitioning function can be remapped freely at every checkpoint, and a heavy key can be split across several partitions, which is what allows close balance even for the heaviest keys.
We also use exact counts merged across subtasks at each checkpoint instead of approximate counters, and apply the resulting function to the next checkpoint window.
Rivetti et al.~\cite{Rivetti:2015} propose distribution-aware key grouping, which learns the frequent keys with a Space Saving sketch, builds an explicit mapping of them and of groups of less frequent keys to partitions, and then deploys it; like \textsc{CheckCo} it relies on frequency estimates taken from the stream, but the mapping is static and does not follow changes in the key distribution, each key is still mapped to a single partition, and the estimates come from approximate counters.
Pacaci and \"Ozsu~\cite{Pacaci:2018:DAP} combine key splitting with an explicit routing table for frequent keys, placed by a greedy heuristic, and hashing for the rest, which keeps the table small.
\textsc{CheckCo} also splits frequent keys, but chooses the number of splits of each key from its estimated count and rebuilds the assignment from the live stream at every checkpoint.

\paragraph{Bin-packing and scheduling}
The LPT algorithm for makespan minimisation on identical machines was analysed by Graham~\cite{Graham:1969:LPT}, who established the $(4/3 - 1/(3P))$-approximation ratio.
We apply LPT to the split-assignment subproblem; the split granularity (Eq.~\eqref{eq:numsplits}) controls how closely the achievable balance approaches the optimum with freely split keys.

\paragraph{Streaming systems}
Apache Flink~\cite{Carbone:2017:Flink} provides the checkpoint barriers we exploit, and its broadcast state~\cite{Flink:BroadcastState} holds the oracle state.
The \texttt{prepareSnapshotPreBarrier} hook lets each subtask emit its local counters before the barrier is forwarded, so the counting adds no work beyond incrementing a counter per event on the hot path.

\paragraph{Frequency estimation over data streams}
Standard approaches for tracking per-key frequencies over unbounded streams include the Count-Min sketch~\cite{Cormode:2005:CMS}, which answers $\varepsilon$-approximate point queries in sublinear space, and the Space Saving algorithm~\cite{Metwally:2005:ECF}, which reports top-$k$ and frequent elements with tight error guarantees.
\textsc{CheckCo}'s oracle avoids sketch-based approximation: each CollectorOperator subtask accumulates exact local counts, which are summed across subtasks at every checkpoint and aggregated in a sliding window of recent checkpoint buckets (Section~\ref{sec:bootstrap}).

\paragraph{Checkpoint-based state management}
The distributed snapshot algorithm of Chandy and Lamport~\cite{Chandy:1985:DS} provides the theoretical basis for consistent global state capture without halting distributed computation.
Carbone et al.~\cite{Carbone:2017:SMA} implement this idea in Apache Flink as lightweight, asynchronous, incremental checkpoints.
\textsc{CheckCo} uses \texttt{prepareSnapshotPreBarrier} to emit local counters before the checkpoint barrier is forwarded, and stores the merged counts in Flink broadcast state, from which the routing table is recomputed on recovery.
Mei et al.~\cite{Mei:2025:Flink2} recently introduce disaggregated state management in Flink~2.0, achieving up to a 94\% reduction in checkpoint duration and 49$\times$ faster recovery after failures or rescaling, relative to Flink~1.20.

\paragraph{Adaptive and learned partitioning}
Several systems have revisited adaptive partitioning since our ICDE 2016 work.
Abdelhamid et al.~\cite{Abdelhamid:2020:Prompt} propose Prompt, which adjusts data partitioning at micro-batch boundaries for Spark-style stream processing systems.
Zapridou et al.~\cite{Zapridou:2022:Dalton} introduce Dalton, a reinforcement-learning partitioner that adapts its policy online and achieves 1.3$\times$--6.7$\times$ higher throughput than existing approaches.
Chen et al.~\cite{Chen:2024:FlexSP} propose FlexSP, a $(1{+}\beta)$-choice partitioning scheme for stateful stream operators in Apache Flink.
Tan et al.~\cite{Tan:2025:SHStream} predict high-frequency keys with a scoring algorithm and split them to reduce skew.
Fragkoulis et al.~\cite{Fragkoulis:2024:Survey} survey load management and elasticity in modern stream processing systems, including skew-aware partitioning, and note that skew mitigation has largely been studied in isolation from other reconfiguration problems.
\textsc{CheckCo} differs from these approaches in that the oracle update is driven entirely by the checkpoint protocol, requiring neither learning nor external coordination, and adds no per-event routing overhead beyond a table lookup and a modulo operation.

%% file: sections/conclusion.tex
\section{Conclusion}
\label{sec:conclusion}

We presented \textsc{CheckCo}, a load-balancing algorithm for skewed event streams that estimates per-key frequencies from the live stream at every Flink checkpoint and gives each key as many workers as its load requires.
\textsc{CheckCo} has run in production for more than a year on a pipeline processing over 100 billion events per day: the most loaded worker receives $9.8\%$ more events than the average (imbalance ratio $1.10$), against $8.03$ for key-based routing in simulation, and the load stays balanced through job restarts.
In simulation, it needs less than half the fan-out of earlier partitioners that reach comparable balance.
Tying the oracle to the checkpoint mechanism removes the need for any external statistics service, as the estimates, their merge and their recovery reuse the machinery that the pipeline already needs for fault tolerance.

\textsc{CheckCo} targets downstream operators without keyed state that must migrate, and its adaptation is tied to the checkpoint interval, so a shift in the key distribution is reflected in routing only from the next checkpoint on.
A natural direction for future work is to decouple the update of the partitioning function from the checkpoint interval.

%% file: references.bib
@String{Computing = "Computing" }

@String{Computer = "{IEEE} Computer" }

@String{Springer = "Springer-Verlag" }

@inproceedings{Breslau:1999:WCZ,
  author    = {Lee Breslau and
               Pei Cao and
               Li Fan and
               Graham Phillips and
               Scott Shenker},
  title     = {Web Caching and Zipf-like Distributions: Evidence and Implications},
  booktitle = {Proceedings of the 18th Annual Joint Conference of the {IEEE} Computer
               and Communications Societies},
  series    = {{INFOCOM} '99},
  volume    = {1},
  pages     = {126--134},
  publisher = {{IEEE}},
  year      = {1999},
  doi       = {10.1109/INFCOM.1999.749260},
}

@article{Graham:1969:LPT,
  author    = {Ronald L. Graham},
  title     = {Bounds on Multiprocessing Timing Anomalies},
  journal   = {{SIAM} Journal on Applied Mathematics},
  volume    = {17},
  number    = {2},
  pages     = {416--429},
  year      = {1969},
  doi       = {10.1137/0117039},
}

@article{Graham:1966:Bounds,
  author    = {Ronald L. Graham},
  title     = {Bounds for Certain Multiprocessing Anomalies},
  journal   = {Bell System Technical Journal},
  volume    = {45},
  number    = {9},
  pages     = {1563--1581},
  year      = {1966},
  doi       = {10.1002/j.1538-7305.1966.tb01709.x},
}

@article{Garey:1978:Strong,
  author    = {M. R. Garey and David S. Johnson},
  title     = {`{Strong}' {NP}-Completeness Results: Motivation, Examples, and Implications},
  journal   = {Journal of the {ACM}},
  volume    = {25},
  number    = {3},
  pages     = {499--508},
  year      = {1978},
  doi       = {10.1145/322077.322090},
}

@article{Gedik:2014,
  author    = {Bu\u{g}ra Gedik},
  title     = {Partitioning Functions for Stateful Data Parallelism in Stream Processing},
  journal   = {The {VLDB} Journal},
  volume    = {23},
  number    = {4},
  pages     = {517--539},
  year      = {2014},
  doi       = {10.1007/s00778-013-0335-9},
}

@inproceedings{Rivetti:2015,
  author    = {Nicol{\`o} Rivetti and
               Leonardo Querzoni and
               Emmanuelle Anceaume and
               Yann Busnel and
               Bruno Sericola},
  title     = {Efficient Key Grouping for Near-Optimal Load Balancing in Stream Processing
               Systems},
  booktitle = {Proceedings of the 9th {ACM} International Conference on Distributed
               Event-Based Systems},
  series    = {{DEBS} '15},
  pages     = {80--91},
  publisher = {{ACM}},
  year      = {2015},
  doi       = {10.1145/2675743.2771827},
}

@article{Carbone:2017:Flink,
  author    = {Paris Carbone and
               Asterios Katsifodimos and
               Stephan Ewen and
               Volker Markl and
               Seif Haridi and
               Kostas Tzoumas},
  title     = {Apache {F}link: Stream and Batch Processing in a Single Engine},
  journal   = {{IEEE} Data Engineering Bulletin},
  volume    = {38},
  number    = {4},
  pages     = {28--38},
  year      = {2015},
}

@inproceedings{Nasir:2015:PBC:ICDE,
  author    = {Muhammad Anis Uddin Nasir and
               Gianmarco De Francisci Morales and
               David Garc\'{\i}a-Soriano and
               Nicolas Kourtellis and
               Marco Serafini},
  title     = {The Power of Both Choices: Practical Load Balancing for Distributed
               Stream Processing Engines},
  booktitle = {Proceedings of the 31st {IEEE} International Conference on Data Engineering},
  series    = {{ICDE} '15},
  pages     = {137--148},
  publisher = {{IEEE}},
  year      = {2015},
  doi       = {10.1109/ICDE.2015.7113279},
}

@inproceedings{Nasir:2016:WTC:ICDE,
  author    = {Muhammad Anis Uddin Nasir and
               Gianmarco De Francisci Morales and
               Nicolas Kourtellis and
               Marco Serafini},
  title     = {When Two Choices Are not Enough: Balancing at Scale in Distributed
               Stream Processing},
  booktitle = {Proceedings of the 32nd {IEEE} International Conference on Data Engineering},
  series    = {{ICDE} '16},
  publisher = {{IEEE}},
  pages     = {589--600},
  year      = {2016},
  doi       = {10.1109/ICDE.2016.7498273},
}

@article{Mitzenmacher:2001:PTW,
  author    = {Michael Mitzenmacher},
  title     = {The Power of Two Choices in Randomized Load Balancing},
  journal   = {{IEEE} Transactions on Parallel and Distributed Systems},
  volume    = {12},
  number    = {10},
  pages     = {1094--1104},
  year      = {2001},
  doi       = {10.1109/71.963420},
}

@article{Cormode:2005:CMS,
  author    = {Graham Cormode and S. Muthukrishnan},
  title     = {An Improved Data Stream Summary: The Count-Min Sketch and its Applications},
  journal   = {Journal of Algorithms},
  volume    = {55},
  number    = {1},
  pages     = {58--75},
  year      = {2005},
  doi       = {10.1016/j.jalgor.2003.12.001},
}

@inproceedings{Metwally:2005:ECF,
  author    = {Ahmed Metwally and Divyakant Agrawal and Amr {El Abbadi}},
  title     = {Efficient Computation of Frequent and Top-$k$ Elements in Data Streams},
  booktitle = {Proceedings of the 10th International Conference on Database Theory},
  series    = {{ICDT} '05},
  pages     = {398--412},
  publisher = {Springer},
  year      = {2005},
  doi       = {10.1007/978-3-540-30570-5_27},
}

@article{Chandy:1985:DS,
  author    = {K. Mani Chandy and Leslie Lamport},
  title     = {Distributed Snapshots: Determining Global States of Distributed Systems},
  journal   = {{ACM} Transactions on Computer Systems},
  volume    = {3},
  number    = {1},
  pages     = {63--75},
  year      = {1985},
  doi       = {10.1145/214451.214456},
}

@article{Carbone:2017:SMA,
  author    = {Paris Carbone and
               Stephan Ewen and
               Gyula F{\'{o}}ra and
               Seif Haridi and
               Stefan Richter and
               Kostas Tzoumas},
  title     = {State Management in {Apache Flink}: Consistent Stateful Distributed Stream Processing},
  journal   = {Proceedings of the {VLDB} Endowment},
  volume    = {10},
  number    = {12},
  pages     = {1718--1729},
  year      = {2017},
  doi       = {10.14778/3137765.3137777},
}

@inproceedings{Abdelhamid:2020:Prompt,
  author    = {Ahmed S. Abdelhamid and
               Ahmed R. Mahmood and
               Anas Daghistani and
               Walid G. Aref},
  title     = {Prompt: Dynamic Data-Partitioning for Distributed Micro-Batch Stream Processing Systems},
  booktitle = {Proceedings of the 2020 {ACM} {SIGMOD} International Conference on Management of Data},
  series    = {{SIGMOD} '20},
  pages     = {2455--2469},
  publisher = {{ACM}},
  year      = {2020},
  doi       = {10.1145/3318464.3389713},
}

@article{Zapridou:2022:Dalton,
  author    = {Eleni Zapridou and
               Ioannis Mytilinis and
               Anastasia Ailamaki},
  title     = {Dalton: Learned Partitioning for Distributed Data Streams},
  journal   = {Proceedings of the {VLDB} Endowment},
  volume    = {16},
  number    = {3},
  pages     = {491--504},
  year      = {2022},
  doi       = {10.14778/3570690.3570699},
}

@inproceedings{Chen:2024:FlexSP,
  author    = {Siyuan Chen and Decheng Zuo and Zhan Zhang},
  title     = {{FlexSP}: $(1 + \beta)$-Choice Based Flexible Stream Partitioning for Stateful Operators},
  booktitle = {Proceedings of the 53rd International Conference on Parallel Processing},
  series    = {{ICPP} '24},
  pages     = {732--741},
  publisher = {{ACM}},
  year      = {2024},
  doi       = {10.1145/3673038.3673157},
}

@article{Fragkoulis:2024:Survey,
  author    = {Marios Fragkoulis and
               Paris Carbone and
               Vasiliki Kalavri and
               Asterios Katsifodimos},
  title     = {A Survey on the Evolution of Stream Processing Systems},
  journal   = {The {VLDB} Journal},
  volume    = {33},
  number    = {2},
  pages     = {507--541},
  year      = {2024},
  doi       = {10.1007/s00778-023-00819-8},
}

@article{Mei:2025:Flink2,
  author    = {Yuan Mei and
               Zhaoqian Lan and
               Lei Huang and
               Yanfei Lei and
               Han Yin and
               Rui Xia and
               Kaitian Hu and
               Paris Carbone and
               Vasiliki Kalavri and
               Feng Wang},
  title     = {Disaggregated State Management in {Apache Flink} 2.0},
  journal   = {Proceedings of the {VLDB} Endowment},
  volume    = {18},
  number    = {12},
  pages     = {4846--4859},
  year      = {2025},
  doi       = {10.14778/3750601.3750609},
}

@misc{Flink:BroadcastState,
  author       = {{The Apache Software Foundation}},
  title        = {The Broadcast State Pattern},
  howpublished = {Apache Flink Documentation},
  year         = {2026},
  note         = {\url{https://nightlies.apache.org/flink/flink-docs-stable/docs/dev/datastream/fault-tolerance/broadcast_state/}}
}

@misc{Iceberg:Spec,
  author       = {{The Apache Software Foundation}},
  title        = {Apache {Iceberg} Table Spec},
  howpublished = {Apache Iceberg Documentation},
  year         = {2026},
  note         = {\url{https://iceberg.apache.org/spec/}}
}

@inproceedings{Dageville:2016:Snowflake,
  author    = {Benoit Dageville and
               Thierry Cruanes and
               Marcin Zukowski and
               Vadim Antonov and
               Artin Avanes and
               Jon Bock and
               Jonathan Claybaugh and
               Daniel Engovatov and
               Martin Hentschel and
               Jiansheng Huang and
               Allison W. Lee and
               Ashish Motivala and
               Abdul Q. Munir and
               Steven Pelley and
               Peter Povinec and
               Greg Rahn and
               Spyridon Triantafyllis and
               Philipp Unterbrunner},
  title     = {The {Snowflake} Elastic Data Warehouse},
  booktitle = {Proceedings of the 2016 International Conference on Management of Data ({SIGMOD})},
  pages     = {215--226},
  year      = {2016},
  doi       = {10.1145/2882903.2903741},
}

@inproceedings{Pacaci:2018:DAP,
  author    = {Anil Pacaci and
               M. Tamer {\"O}zsu},
  title     = {Distribution-Aware Stream Partitioning for Distributed Stream Processing Systems},
  booktitle = {Proceedings of the 5th {ACM} {SIGMOD} Workshop on Algorithms and Systems for {MapReduce} and Beyond ({BeyondMR})},
  year      = {2018},
  doi       = {10.1145/3206333.3206338}
}

@article{Liu:2024:FlexD,
  author    = {Gang Liu and
               Zeting Wang and
               Amelie Chi Zhou and
               Rui Mao},
  title     = {Adaptive Key Partitioning in Distributed Stream Processing},
  journal   = {{CCF} Transactions on High Performance Computing},
  volume    = {6},
  number    = {2},
  pages     = {164--178},
  year      = {2024},
  doi       = {10.1007/s42514-023-00179-3}
}

@article{Tan:2025:SHStream,
  author    = {Jiawei Tan and
               Li Yang and
               Yaolian Guo and
               Zhiwei Zuo and
               Xiong Xiao},
  title     = {A Distributed Skewed Stream Processing System Based on Scoring High-Frequency Key Perception},
  journal   = {The Journal of Supercomputing},
  volume    = {81},
  number    = {9},
  pages     = {1073},
  year      = {2025},
  doi       = {10.1007/s11227-025-07465-7}
}

@article{Azar:1999:BA,
  author    = {Yossi Azar and Andrei Z. Broder and Anna R. Karlin and Eli Upfal},
  title     = {Balanced Allocations},
  journal   = {{SIAM} Journal on Computing},
  volume    = {29},
  number    = {1},
  pages     = {180--200},
  year      = {1999},
  doi       = {10.1137/S0097539795288490},
}
